\documentstyle[
               12pt,
               prd,aps,epsfig]{revtex}

\newenvironment{equation*}{\begin{displaymath}}{\end{displaymath}}
\newcommand{\const}{\mbox{const} }
\newcommand{\Scri}{\mbox{$\cal J$}}

\newcommand{\N}[1]{{\cal N}_{\rm\bf #1\>}{}}

\newcommand{\DIII}{\,{}^{\scriptscriptstyle(3)\!\!\!\:}\nabla}
\newcommand{\DeIII}{\,{}^{\scriptscriptstyle(3)\!\!\!\:}\Delta}

\newcommand{\ROI}{\,{}^{\scriptscriptstyle(0,1)\!\!\!\:}\hat R}
\newcommand{\RII}{\,{}^{\scriptscriptstyle(1,1)\!\!\!\:}\hat R}
\newcommand{\epsIII}{\,{}^{\scriptscriptstyle(3)\!\!\!\:}\epsilon}
\def\@warning#1{\typeout{LaTeX Warning [l.\the\inputlineno]: #1.}}

\begin{document}


\title{A Scheme to Numerically Evolve Data for the Conformal Einstein Equation}

\author{{\bf Peter H\"ubner}\\
        (pth@aei-potsdam.mpg.de)\\
        Max-Planck-Institut f\"ur Gravitationsphysik\\
        Albert-Einstein-Institut\\
        Schlaatzweg 1\\
        D-14473 Potsdam\\
        FRG}

\maketitle
\thispagestyle{empty}
\mbox{}
\\
{\small short title: Integrating the Conformal Einstein Equation}
\\
\mbox{}

\begin{abstract}
This is the second paper in a series describing a numerical
implementation of the conformal Einstein equation.
This paper deals with the technical details of the numerical code used
to perform numerical time evolutions from a ``minimal'' set of data.
\\
We outline the numerical construction of a complete set of data for our
equations from a minimal set of data.
The second and the fourth order discretisations, which are used for the 
construction of the complete data set and for the numerical
integration of the time evolution equations, are described and their
efficiencies are compared.
By using the fourth order scheme we reduce our computer resource
requirements --- with respect to memory as well as computation time
--- by at least two orders of magnitude as compared to the second
order scheme.
\end{abstract}


%
%
\section{Introduction}
We calculate solutions to the Einstein equation arising from
hyperboloidal initial data by solving the conformal field equation.
In the first paper in this series~\cite{Hu98bh} the ideas behind
the conformal approach, their mathematical foundation, and the
benefits of it have been discussed in detail. 
In this paper we discuss the numerical details of the part of the
implementation which calculates the time evolution from a minimal set
of data consisting of the conformal 3-metric $h_{ab}$ and the
conformal extrinsic curvature $k_{ab}$ as well as the conformal factor
$\Omega$ and its time derivative $\Omega_0$.
The latter relate $(h_{ab},k_{ab})$ to the physical data.
\\
To be able to test the numerics described here, we take known exact
solutions given in terms of the conformal metric $g_{ab}$ and
the conformal factor $\Omega$, perform numerical coordinate
transformations to hide obvious symmetries, and calculate in a
straightforward way a minimal set of data from the transformed
solution.
Then we extend the minimal set of data to a complete set of conformal
data.
The calculation of the complete data set is a delicate numerical
issue.
In section~\ref{VolleDaten} we discuss why it is so and we show how to
solve this problem.
\\
After having calculated a complete set of data we can start the actual
time integration.
In section~\ref{discretisation} we describe in detail the second and
the fourth order scheme used to numerically integrate the symmetric
hyperbolic time evolution equations.
\\
In the section~\ref{Tests} we outline properties of the exact solutions
important for the tests, the tests themselves, and the results of the
comparisons between the two schemes.
\\
In the tests based on conformal Minkowski data  we reconstruct the
whole future of the initial slice as well as null infinity and even
timelike infinity after a finite number of time steps. 
Obviously we can then stop the calculation.
As suggested by the theorems in~\cite{Fr91ot} we can expect to cover the future
of the initial slice in all cases of sufficiently small asymptotically
Minkowskian data, i.~e.\ for all gravitational wave data not forming
singularities or black holes, by one finite grid.
\\
In the other tests we use asymptotically A3 solutions
(cf.\ subsection~\ref{ExactLsg}).
These are solutions which in general contain gravitational radiation
but are special in the sense that they possess a conformal structure
which becomes singular towards timelike infinity.
We compare the numerical result with the analytic solution after
covering a certain integration time.
To check stability we have also performed much larger numbers of
time steps and we did not encounter any numerical problems, but the
numerical evolution slows down for the following reason.
In the given coordinates, the light cones in the interior become
flatter and flatter the closer they are to the singular timelike 
infinity.
Therefore, the  Courant-Friedrichs-Levy condition must enforce smaller
and smaller time steps and it prevents us from reaching timelike 
infinity in the numerical integration, unless we embark upon a more
detailed study (which is beyond the goal of this paper).
\\
Comparing the second order with the fourth order scheme we found the
fourth order scheme surpassing the second order scheme in three
dimensional calculations by an at least two orders of magnitude more
efficient use of computer resources.
In all cases we tested the combination of the fourth order scheme with
grids with $100$ gridpoints in each space dimensions was sufficient to
achieve relative errors of less than one percent. 
The resulting moderate memory and computer time requirements allow us
to do our calculations, including the ones with three space dimensions,
on medium size parallel computers which are nowadays available in many
physics institutes with numerical orientation.
\\
In appendix~\ref{ComScieAsp} we give a short review of the
computational science aspects of the code.
%
%

%
%
\section{Constructing a complete data set}
\label{VolleDaten}
To avoid too many repetitions, we assume the reader to be familiar
with the general approach as well as with the equations discussed in
part one of this series, and we shall refer to equation~(n) of part
one by writing~(I/n).
\subsection{Minimal set of data from exact solutions}
The known solutions of the conformal field equations have all high
degrees of symmetries, and usually the metric $g_{a'b'}$ and the
rescaling factor $\Omega$ are written in terms of symmetry adapted
coordinates $x_s{}^{a'} = (t_s,x_s,y_s,z_s)$.
In these coordinates many variables and major parts
of the equations are identically zero, hence they are of limited use
for testing.
Things become more interesting if we make coordinate transformations
which hide the symmetries.
For technical reason we prescribe the inverse of the coordinate
transformation,
\begin{equation}
\label{KoordTrans}
  x_s{}^{a'} \> = \> x_s{}^{a'}(x^a).
\end{equation}
Then the components of the metric $g_{ab}$ in the new coordinates
$g_{ab}$ are
\begin{equation}
\label{MetrikTrans}
  g_{ab}(x^a) = 
    g_{a'b'}(x_s{}^{a'}(x^a)) 
      (\partial_a x_s{}^{a'}) (\partial_b x_s{}^{b'}).
\end{equation}
The derivatives are numerically calculated by evaluating fourth order
stencils according to~(\ref{AblDiskrIV}) with respect to an imaginary
grid with one tenth of the grid spacing of the real grid.
From equation~(\ref{MetrikTrans}) we can read off the first elements of
a minimal set of data on the hypersurface defined by $t=t_0$, namely the
components of the 3-metric $h_{ab}$, 
as the \{$x$,$y$,$z$\}-components of $g_{ab}$.
The next element of a minimal set, the scalar $\Omega$, only changes
its functional form under the coordinate transformation,
$\Omega(x^a)=\Omega(x_s{}^{a'}(x^a))$.
\\
After numerically calculating time and space derivatives of
($h_{ab}$,$\Omega$), equation~(I/13a) respectively~(I/13h) are used to
calculate the missing elements ($k_{ab}$,$\Omega_0$) of a minimal set
of data.
The needed lapse $N$ and shift $N^a$ are obtained as the solution of the
system of equations $g_{tt} = - N^2 + h_{ab} N^a N^b$ and $g_{\{x,y,z\}t} =
h_{\{x,y,z\}b} N^b$.
\\
During the time evolution we have to prescribe the gauge source
functions $q=\ln(N/\sqrt{h})$, shift $N^a$, and Ricci scalar
$R$.
The function $q=\ln(N/\sqrt{h})$ and the shift $N^a$ are
calculated from the coefficients of $g_{ab}$ on each slice in analogy
to the approach on the initial slice, the Ricci scalar $R$ is given as
function of the coordinates.
\subsection{From a minimal to a complete set of conformal data}
To calculate the remaining initial data $(\gamma^a{}_{bc}$, $\ROI_a$,
$\RII_{ab}$, $E_{ab}$, $B_{ab}$, $\Omega_a$, $\omega$) from the
minimal set we use certain combinations of the conformal constraints
(I/14).
\begin{mathletters}
In particular, we use 
\begin{eqnarray}
  \N{h}_{abc} & = & 0,\\
  \N{k}_{ab}{}^b & = & 0,\\
  \N{\gamma}_{ab}{}^{ab} & = & 0,\\
  \N{\Omega}_{a} & = & 0,\\
  \N{\Omega_a}_{a}{}^{a} & = & 0,\\
  \N{\Omega_a}_{ab} & = & 0,\\
  \epsIII_a{}^{cd}\N{k}_{cdb} & = & 0,
\end{eqnarray}
to solve for $\gamma^a{}_{bc}$, $\ROI_a$, $\RII_a{}^a$, $\Omega_a$,
$\omega$, $f_{\RII\,ab} := \Omega \RII_{ab}$, $f_{B\,ab}:= \Omega
B_{ab}$ respectively.
\\
To extract $\RII_{ab}$ and $B_{ab}$ from $f_{\RII\,ab}$ and
$f_{B\,ab}$ we have to divide by $\Omega$.
This division by $\Omega$ is numerically the most delicate part of the
construction of a complete set of data.
It will be described in the next subsection.
After the division we use $\RII_{ab}$ to calculate $f_{E\,ab} :=
\Omega E_{ab}$ from
\begin{eqnarray}
  \N{\gamma}_{acb}{}^{c} & = & 0.
\end{eqnarray}
To determine $E_{ab}$ from $f_{E\,ab}$ we again have to divide by
$\Omega$.
\end{mathletters}
\subsection{Dividing by $\Omega$}
\label{DivdurchOm}
\begin{figure}[htbp]
  \begin{center}
      \begin{minipage}[t]{5.1cm}
        \input durchOm.pstex_t
        \vskip0.5em
        \caption{\label{durchOm}Sketch of $\RII_{ab}$ (solid line)
          and $E_{ab}$ (dashed line) near ${\cal S}$ as obtained by a
          combination of a division by $\Omega$ outside ${\cal S}$ and
          the application of l'Hopital's rule at ${\cal S}$. The
          crosses denote gridpoints.}
      \end{minipage}
    \hspace{2em}
      \begin{minipage}[t]{9.1cm}
        \epsfxsize=6.5cm
        \centerline{\epsfbox{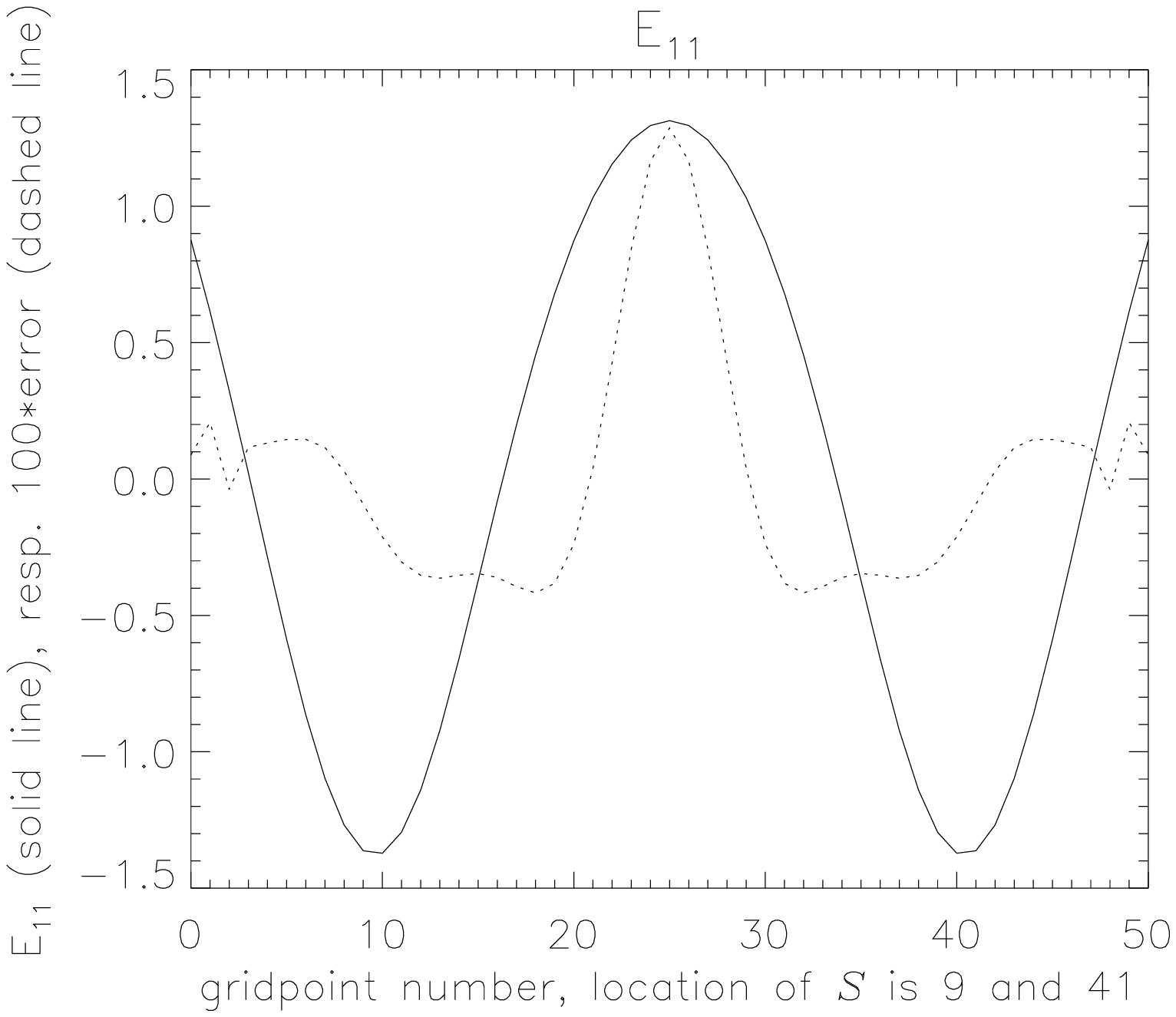}}
        \vskip0.5em
        \caption{\label{durchOmEll}Result of our method applied to an
          A3 spacetime on a $50^3$ grid. The solid line shows
          $E_{11}+2.5$ along a typical $(y,z)=\const$ line, which
          intersects ${\cal S}$ at gridpoints 9 and 41. The dashed line
          shows 100 $\times$ the difference to an $100^3$ calculation. Its
          non-smoothness near the boundary is caused by changing from a
          symmetric to an asymmetric stencil. It does not affect the
          boundary treatment described in subsection~\ref{Randbehandlung}.}  
      \end{minipage}
  \end{center}
\end{figure}

In~\cite{AnCA92ot} the smoothness of the limits
$\lim_{\Omega\rightarrow 0} \frac{f}{\Omega}$ has been analysed in
detail.
In a straightforward implementation of the approach used there we
would divide by $\Omega$ outside the set ${\cal S}$ on the initial
slice and use l'Hopital's rule at ${\cal S}$, where $\Omega$ vanishes.
Then a picture like the one shown in figure~\ref{durchOm} would
result.
The reason for the pole like structure near ${\cal S}$ is simple, as we
will immediately see. 
Let us assume, for simplicity, that we have one space dimension only,
that $\Omega=0$ at $x=0$, and that the corresponding gridpoint has
number $0$.
In the process of calculating ($f_{\RII}$, $f_{B}$,$f_{E}$) numerical
derivatives are taken, therefore each $f$ deviates from the exact
value $f_e$ by a discretisation error, $f=f_e+a_f (\Delta x)^n$, where
$n$ is the order of the scheme.
We now make a Taylor expansion of $f_e$ and $\Omega$ around $x=0+dx$ and
recognise that we lost near point $0$ one order of accuracy, since
$\frac{f}{\Omega}=\frac{f_e}{\Omega} + \frac{a_f}{d_x\Omega|_0}
\frac{\Delta x}{dx} \, (\Delta x)^{n-1}$ and $\frac{\Delta x}{dx}$ is
always of order $1$ at the neighbours of gridpoint $0$.
$\frac{\Delta x}{dx}$ changes sign at point $0$, hence the form of the
solid line in figure~\ref{durchOm}.
When dividing by $\Omega$ a second time to calculate $E_{ab}$ we see
by the same kind of argument and by recalling, that the pole like
structure of the solid line enters into $f_{E\,ab}$, that we get the
dashed line in figure~\ref{durchOm}.
\\
This non-smoothness of the discretisation error and the loss of two
orders of accuracy in the initial data is unacceptable mainly for two
reasons.
Firstly, discretisation schemes for symmetric hyperbolic equations
tend to react to these kinds of non-smoothness in the initial
data with significant drops in the convergence order.
Secondly, this defect would be most significant in the neighbourhood
of ${\cal S}$ and would hence establish a kind of boundary at ${\cal
  S}$ which would eliminate many advantages of the conformal method. 
\\
We tried various methods to remove the pole-like structure by some
kind of smoothing procedure.
Although the behaviour could be improved, the phenomena never
disappeared completely.
The problematic behaviour could almost be cured by first subtracting a
function with values of the order of the discretisation error from $f$
on the whole grid, making $f$ exactly vanish at ${\cal S}$, and then
dividing by $\Omega$.\footnote{J\"org Frauendiener has developed a
  sophisticated way to combine this kind of idea with spectral
  decomposition in his 2D code for asymptotically A3
  spacetimes~\cite{Fr98ci}.}
We did not pursue this approach beyond the asymptotically A3
scenarios because we did not see how to generalise it to arbitrary ${\cal
  S}$ and in particular because we found the following approach which
works for arbitrary ${\cal S}$.
\\
Instead of solving the equation $g \Omega =f$ for $g$, we determine
$g$ as solution to an elliptic equation of the type
\begin{equation}
\label{ellFdurchHstart}
  \DeIII ( \Omega^2 g - \Omega f ) = 0, 
\end{equation}
where the symbol $\DeIII$ denotes the Laplace operator of $\DIII_a$,
that is $\DeIII := \DIII^a \DIII_a$.
If we write this as an equation for $u:=\Omega^2 g - \Omega f$, we see
that the  uniquely determined solution for the boundary values
$u\mid_{\rm   boundary}=0$ is $u=0$, i.~e.\ $g=f/\Omega$.
\\
Written as equation for $g$ equation (\ref{ellFdurchHstart}) reads
\begin{eqnarray}
\label{ellFdurchHexp}
  \lefteqn{
    \Omega^2 \, \DeIII g 
    \, + \, 4 \Omega \left(\DIII^a \Omega\right) \, \left(\DIII_a g\right)
    \, + \, \left[ 2 \left(\DIII^a \Omega\right) \left(\DIII_a \Omega\right)
                   + 2 \Omega \DeIII \Omega \right] \, g \> = }\hspace{18em}
 \nonumber\\
 & & f \DeIII \Omega + 2 \left(\DIII^a \Omega\right) \left(\DIII_a f\right)
     + \Omega \DeIII f.
\end{eqnarray}
If the derivatives are discretised by symmetric stencils
like~(\ref{DiskrII}) or~(\ref{DiskrIV}), the $\left(\DIII_a g\right)$
term has a sign which is known to cause a tendency for
instabilities~\cite{St98pc}.
To avoid the resulting numerical problems we add
\begin{equation}
  \label{ellFdurchHkorr}
  - \eta \left( \DIII^a \Omega \right) 
    \left( \DIII_a ( \Omega g - f ) \right) = 0
\end{equation}
to equation~(\ref{ellFdurchHexp}) and obtain
\begin{eqnarray}
\label{ellFdurchH}
  \lefteqn{
    \Omega^2 \, \DeIII g 
    \, + \, (4-\eta) \Omega \left(\DIII^a \Omega\right) 
         \,  \left(\DIII_a g\right)
    \, + \, \left[ (2-\eta) \left(\DIII^a \Omega\right) 
            \left(\DIII_a \Omega\right)
       + 2 \Omega \DeIII \Omega \right] \, g \> = }\hspace{16em}
 \nonumber\\
  & & f \DeIII \Omega 
      + ( 2 - \eta ) \left(\DIII^a \Omega\right) \left(\DIII_a f\right)
      + \Omega \DeIII f.
\end{eqnarray}
This is the linear elliptic equation we solve after we have
numerically calculated the function $f$, which is due to
discretisation errors not exactly $0$ at ${\cal S}$.
Instead of having to divide by $\Omega$ on the whole grid we only have 
to divide by $\Omega$ on the true\footnote{For a definition of
  ``true'' in the context of boundaries see
  subsection~\ref{Randbehandlung}.} grid boundaries to provide boundary
values for our elliptic equation.
This does not constitute any numerical problem, since $\Omega$ is
significantly different from $0$ at the true grid boundaries.
\\ 
Equation~(\ref{ellFdurchH}) tells us that on ${\cal S}$ 
\begin{equation}
\label{LimitOnS}
g = \frac{f \DeIII \Omega}{( 2 - \eta )\left(\DIII^a
    \Omega\right)\left(\DIII_a \Omega\right)} 
    + \frac{\left(\DIII^a \Omega\right) \left(\DIII_a
        f\right)}{\left(\DIII^a \Omega\right)\left(\DIII_a
        \Omega\right)},
\end{equation}
which is a correct answer, since $f$ vanishes up to the discretisation
error.
Since $\DIII_a\Omega\mid_{\cal S}\ne0$ in any extended hyperboloidal
initial value problem, expression~(\ref{LimitOnS}) is well-defined.
\\
Of course the properties of equation~(\ref{ellFdurchH}) depend on the
parameter $\eta$:
For $\eta=0$ it is difficult to get a stable numerical scheme, for
$\eta=2$ there is no solution possible if $f\mid_{\cal S}\ne 0$, for
$\eta=3$ the discretised system can be written as symmetric
matrix~\cite[equation(5.1.18)]{Ha96TU}, for $\eta=4$ the term
$\left(\DIII_a g\right)$ has vanishing coefficient, and for $\eta=8$ 
the system possesses the same principal part as the Yamabe equation,
which will play a crucial role in the next part of the series, where
we describe how to generate minimal sets of data not representing
known exact solutions.
At least the later three choices for $\eta$ deliver values of $g$ which converge
to the exact $g$ with a smooth discretisation error and without loss
of one order of convergence.
Figure~\ref{durchOmEll} shows the smoothness and the error for a run
with a pretty coarse grid.
\\
The default choice in the code is $\eta=8$, together with
$\delta_{ab}$ as 3-metric.
To exclude the possibility of having calculated a spurious solution of 
equation~(\ref{ellFdurchH}), except for the $\eta=0$ case we have
not shown uniqueness, we evaluate the constraints (I/14) to check
consistency.
\\[0.3em]
To discretise we substitute
\begin{mathletters}
\label{DiskrII}
\begin{eqnarray}
\label{AblDiskrII}
  \partial_x f  & \rightarrow &
    \frac{1}{2 \Delta x} \left( f_{i+1,j,k} - f_{i-1,j,k} \right),
\\
\label{EllDiskrII}
  \partial_x{}^2 f & \rightarrow &
    \frac{1}{(\Delta x)^2}
    \left( f_{i,j,k} - 2 f_{i,j,k} + f_{i-1,j,k} \right),
\end{eqnarray}
\end{mathletters}
and their equivalents for the $y$ and $z$ coordinates to obtain second
order approximations, respectively 
\begin{mathletters}
\label{DiskrIV}
\begin{eqnarray}
\label{AblDiskrIV}
  \partial_x f & \rightarrow &
    \frac{1}{12 \Delta x} 
    \left( - f_{i+2,j,k} + 8 f_{i+1,j,k} - 8 f_{i-1,j,k} + f_{i-2,j,k}
    \right),
\\
\label{EllDiskrIV}
  \partial_x{}^2 f & \rightarrow &
    \frac{1}{12 (\Delta x)^2}
    \left( - f_{i+2,j,k} + 16 f_{i+1,j,k} - 30 f_{i,j,k}
           + 16 f_{i-1,j,k} - f_{i-2,j,k} \right),
\end{eqnarray}
\end{mathletters}
and their equivalents for the $y$ and $z$ coordinates to obtain fourth
order approximations.
\\
The elliptic equation~(\ref{ellFdurchH}) then becomes a set of linear
equations for the values $g_{i,j,k}$ of $g$ at gridpoints $(i,j,k)$,
denoted by the vector $\underline{u}$:
\begin{equation}
\label{ellDiskrGl}
  \underline{\underline{C}} \, \underline{u} = \underline{r},
\end{equation}
with some sparse matrix $\underline{\underline{C}}$ depending on the
stencils.
The vector $\underline{r}$ is given by the prescribed boundary values
and the right hand side of equation~(\ref{ellFdurchH}).
The matrix $\underline{\underline{C}}$ would become singular if
$\Omega$ and $\partial_a \Omega$ vanished simultaneously at one
gridpoint.
For an extended hyperboloidal initial value problem this cannot happen.
\\
Equation~(\ref{ellDiskrGl}) is solved iteratively by using the
algebraic multigrid library (AMG) by K. St\"uben from the Gesellschaft
f\"ur Mathematik und Datenverarbeitung.
AMG analyses the algebraic structure of the matrix
$\underline{\underline{C}}$ and derives from the structure a strategy 
to apply multigrid techniques to accelerate the convergence rate of
the iterative solution of equation~(\ref{ellDiskrGl}).
Since the structure analysis happens automatically in AMG, it is
very easy to program elliptic solvers for different stencils or for
grids with different topologies, one only needs to change the
computation of $\underline{\underline{C}}$ and $\underline{r}$, but
not the multigrid part.
This convenience more than outweighs the computational overhead from
analysing the algebraic structure of $\underline{\underline{C}}$ once
in each program run.
The interested reader can find a detailed description of AMG
in~\cite{St99am,RuS87am}.
\\
Before we end this section, we should make a remark about the accuracy
which we can achieve.
Increasing the number of gridpoints decreases the discretisation
error.
On the other side, it is well-known that matrix inversion, here
solving equation~(\ref{ellDiskrGl}), amplifies rounding errors.
The amplification grows with the size of the matrix
$\underline{\underline{C}}$, which is determined by the number of
gridpoints.
Due to the amplification of the rounding errors refinement of
the grid beyond a certain threshold will not improve the accuracy of
the solution.
When we surpass the $500^2$ grid size in a fourth order scheme
calculation, rounding errors become a visible contribution to the
total error.
The size of the relative error in the solutions of the elliptic solver
is then of order $10^{-11}$.
A remnant of that lower bound to the convergence of the scheme becomes
visible in the curve for the $640^2$ grid size in
figure~\ref{IVteA3KonvConstr}.
%
%
%

%
%
\section{The discretisation of the time integrator}
\label{discretisation}
In this section, where we describe the discretisation of the time
evolution equations~(I/13), we write our equations formally as
\begin{equation}
\label{GleichungFormal}
  \partial_t \underline{f} 
  + \underline{\underline{A}}^i(f) \partial_i \underline{f}
  = \underline{b}(\underline{f}).
\end{equation}
Our system of equation is, up to some simple algebraic manipulations,
a quasilinear symmetric hyperbolic system of first order.
One of the characteristic properties of these kinds of systems is the
existence of a maximal set of real characteristics and thus of a
finite propagation speed of signals.
In implicit schemes the numerical speed of propagation is infinite, or 
at least very large.
To mimic the finiteness of the propagation speed we therefore have to
use explicit schemes.
Moreover, we have to expect that parts of our slices run into
singularities during time evolution.
Due to the infinite numerical propagation speed in implicit schemes,
the occurrence of a single gridpoint with singular values of a single
variable would cease the calculation.
In explicit schemes we can in principle continue the calculation to cover at
least part of the remaining spacetime~\cite{CoS83nr,Hu96na}.
For these reasons we have only evaluated explicit schemes.
\subsection{The second order scheme}
\subsubsection{General considerations for the choice of the scheme}
The general form of explicit second order schemes in 1D\footnote{We
  call a calculation nD, if it uses an n-dimensional space grid.}  is
given by the so-called $S^\alpha_\beta$ schemes~\cite{PeT83CM}, which
contain the widely used MacCormack and Lax-Wendroff schemes.
Their extension to 2D and 3D is not unique and most of the schemes
obtained distinguish certain propagation directions.
For many of the schemes with a distinguished direction, e.~g.\ all
generalisations of the MacCormack scheme, the occurrence of a weak
instability could be shown already for the advection equation, if the
propagation direction is opposite to the distinguished
direction~\cite{Tu74pe}.
This instability is of a strange character and hard to 
detect~\cite{Tu74pe,RiM67DM}.
Since for quasilinear equations the propagation directions depend on
the data and may change during the time evolution, those schemes, even
if initially stable, may become unstable.
\\
To avoid distinguishing certain directions and dealing with
these instabilities we implemented the rotated Richtmyer scheme, the
extension of Lax-Wendroff which does not distinguish
propagation directions.
For our evolution equations combined with strong data even the rotated
Richtmyer scheme turned out to be unstable.
A grid mode with a wave length of ten gridpoints was not damped
sufficiently.
Although we could make this grid mode vanish by adding artificial
viscosity, the scheme was still not stable, a 20 gridpoint grid mode
appeared later in the time evolution. 
In the linearised equations the growth rate of the instability was
significantly weaker, the linearised equations without source were
stable, as  predicted by the theory.
To be able to treat the principle part, which determines the
propagation directions, by rotated Richtmyer and the sources by
something else, we use Strang splitting.
\subsubsection{The implemented second order scheme}
In the  Strang splitting ansatz one formally writes
equation~(\ref{GleichungFormal}) as an equation for the principal part,
\begin{equation}
\label{HauptteilFormal}
  \partial_t \underline{f} 
  + \underline{\underline{A}}^i(f) \partial_i \underline{f}
  = 0,
\end{equation}
and an equation for the sources,
\begin{equation}
\label{QuelleFormal}
  \partial_t \underline{f} 
  = \underline{b}(f).
\end{equation}
To integrate equation~(\ref{HauptteilFormal}) we use the rotated Richtmyer
scheme:
\\
First, we calculate a half grid,
\begin{eqnarray}
\label{rRHauptteil}
  \underline{f}^{l}_{i+1/2,j+1/2,k+1/2} & = &
    \frac{1}{2^3} 
      \left(   \underline{f}^{l}_{i    ,j    ,k    }
             + \underline{f}^{l}_{i+1/2,j    ,k    }
             + \underline{f}^{l}_{i    ,j+1/2,k    }
             + \underline{f}^{l}_{i+1/2,j+1/2,k    } \right.
\nonumber \\
  & & \hphantom{\frac{1}{2^3} \hspace{.5em}} \left.
             + \underline{f}^{l}_{i    ,j    ,k+1/2}
             + \underline{f}^{l}_{i+1/2,j    ,k+1/2}
             + \underline{f}^{l}_{i    ,j+1/2,k+1/2}
             + \underline{f}^{l}_{i+1/2,j+1/2,k+1/2} \right),
\nonumber \\
\noalign{and the derivatives thereon,}
  \partial_x \underline{f}^{l}_{i+1/2,j+1/2,k+1/2} & = &
    \frac{1}{2^{3-1} \Delta x} 
      \left( - \underline{f}^{l}_{i  ,j  ,k  }
             + \underline{f}^{l}_{i+1,j  ,k  }
             - \underline{f}^{l}_{i  ,j+1,k  }
             + \underline{f}^{l}_{i+1,j+1,k  } \right.
\nonumber \\
  & & \hphantom{\frac{1}{2^{3-1} \Delta x} \hspace{.5em}} \left.
             - \underline{f}^{l}_{i  ,j  ,k+1}
             + \underline{f}^{l}_{i+1,j  ,k+1}
             - \underline{f}^{l}_{i  ,j+1,k+1}
             + \underline{f}^{l}_{i+1,j+1,k+1} \right), 
\nonumber \\
\noalign{$\partial_y \underline{f}^{l}_{i+1/2,j+1/2,k+1/2}$ and
  $\partial_z \underline{f}^{l}_{i+1/2,j+1/2,k+1/2}$ in analogy,}
\nonumber
\end{eqnarray}
to take the predictor step:
\begin{equation*}
  \underline{f}_{\cal P}{}^{l+1/2}_{i+1/2,j+1/2,k+1/2} =
    \underline{f}^{l}_{i+1/2,j+1/2,k+1/2} 
    - \frac{\Delta t}{2}
      \underline{\underline{A}}^m(\underline{f}^{l}_{i+1/2,j+1/2,k+1/2})
        \partial_m \underline{f}^{l}_{i+1/2,j+1/2,k+1/2}.
\end{equation*}
Then, we average again, 
\begin{eqnarray}
  \underline{f}_{\cal P}{}^{l+1/2}_{i,j,k} & = &
    \frac{1}{2^3} 
      \left(   \underline{f}_{\cal P}{}^{l}_{i-1/2,j-1/2,k-1/2}
             + \underline{f}_{\cal P}{}^{l}_{i+1/2,j-1/2,k-1/2}
             + \underline{f}_{\cal P}{}^{l}_{i-1/2,j+1/2,k-1/2} \right.
\nonumber \\
  & & \hphantom{\frac{1}{2^3} \hspace{.5em}} \left.
             + \underline{f}_{\cal P}{}^{l}_{i+1/2,j+1/2,k-1/2}
             + \underline{f}_{\cal P}{}^{l}_{i-1/2,j-1/2,k+1/2}
             + \underline{f}_{\cal P}{}^{l}_{i+1/2,j-1/2,k+1/2} \right.
\nonumber \\
  & & \hphantom{\frac{1}{2^3} \hspace{.5em}} \left.
             + \underline{f}_{\cal P}{}^{l}_{i-1/2,j+1/2,k+1/2}
             + \underline{f}_{\cal P}{}^{l}_{i+1/2,j+1/2,k+1/2} \right),
\nonumber \\
\noalign{and again calculate the derivatives,}
  \partial_x \underline{f}_{\cal P}{}^{l+1/2}_{i,j,k} & = &
    \frac{1}{2^{3-1} \Delta x} 
      \left( - \underline{f}_{\cal P}{}^{l}_{i-1/2,j-1/2,k-1/2}
             + \underline{f}_{\cal P}{}^{l}_{i+1/2,j-1/2,k-1/2} \right.
\nonumber \\
  & & \hphantom{\frac{1}{2^{3-1} \Delta x} \hspace{.5em}} \left.
             - \underline{f}_{\cal P}{}^{l}_{i-1/2,j+1/2,k-1/2}
             + \underline{f}_{\cal P}{}^{l}_{i+1/2,j+1/2,k-1/2}
           \right.
\nonumber \\
  & & \hphantom{\frac{1}{2^{3-1} \Delta x} \hspace{.5em}} \left.
             - \underline{f}_{\cal P}{}^{l}_{i-1/2,j-1/2,k+1/2}
             + \underline{f}_{\cal P}{}^{l}_{i+1/2,j-1/2,k+1/2} \right.
\nonumber \\
  & & \hphantom{\frac{1}{2^{3-1} \Delta x} \hspace{.5em}} \left.
             - \underline{f}_{\cal P}{}^{l}_{i-1/2,j+1/2,k+1/2}
             + \underline{f}_{\cal P}{}^{l}_{i+1/2,j+1/2,k+1/2} \right),
\nonumber \\
\noalign{$\partial_y \underline{f}^{l+1/2}_{i,j,k}$ and
  $\partial_z \underline{f}^{l+1/2}_{i,j,k}$ in analogy,}
\nonumber
\end{eqnarray}
to take the corrector step:
\begin{equation}
  \underline{f}_{\cal P}{}^{l+1}_{i,j,k} 
    = {\cal P} \underline{f}^{l}_{i,j,k} 
  :=
    \underline{f}^{l}_{i,j,k} 
    - \Delta t
      \underline{\underline{A}}^m(\underline{f}_{\cal P}{}^{l+1/2}_{i,j,k})
        \partial_m \underline{f}_{\cal P}{}^{l+1/2}_{i,j,k}.
\end{equation}
To integrate the source equation~(\ref{QuelleFormal}) we use the
pseudo-implicit Heun scheme~\cite{EnR84FZ}, since it is said to be
similarly robust in the case of stiff equations as implicit schemes:
\begin{eqnarray}
\label{HeunQuelle}
  \underline{f}_{\cal S}{}^{l+1/n}_{i,j,k} & = &
    \underline{f}_{\cal S}{}^{l}_{i,j,k} 
    + \Delta t \underline{b}(\underline{f}^{l}_{i,j,k})
\nonumber \\
  \underline{f}_{\cal S}{}^{l+m/n}_{i,j,k} & = &
    \underline{f}_{\cal S}{}^{l}_{i,j,k} 
    + \frac{\Delta t}{2}
      \left( \underline{b}(\underline{f}_{\cal S}{}^{l}_{i,j,k})
             + \underline{b}(\underline{f}_{\cal
               S}{}^{l+(m-1)/n}_{i,j,k})
      \right), 
    \qquad 2\le m\le n.
\nonumber     
\end{eqnarray}
Here $n$ is the number of iterations.
\begin{equation}
  \underline{f}_{\cal S}{}^{l+1}_{i,j,k} 
     =: {\cal S} \underline{f}_{\cal S}{}^{l}_{i,j,k} 
\end{equation}
There are two standard ways to combine the integration operator ${\cal
  P}$ for the principal part with the integration operator ${\cal S}$
for the source terms, the Strang I and the Strang II
scheme~\cite{Wi72so}.
They have the disadvantage of consuming additional memory and requiring
many loops over the grids.
By using different schemes for the odd and even time steps, namely
\begin{eqnarray}
  \underline{f}^{l+2 n - 1}_{i,j,k} & = &
    {\cal S} {\cal P} \underline{f}^{l+2 n - 2}_{i,j,k}
\nonumber \\
\noalign{and}
  \underline{f}^{l+2 n}_{i,j,k} & = &
    {\cal P} {\cal S} \underline{f}^{l+2 n - 1}_{i,j,k},
\end{eqnarray}
we avoid these disadvantages and obtain a scheme which is globally
second order, although the coefficient of the leading term in the
discretisation error jumps between odd and even steps.
\\
In the numerical implementation we use three grids to store the
variables, the gauge source functions, and the intermediate values. 
Even if we calculated the gauge source functions by some global
second order procedure, which were of course incompatible with
hyperbolicity of the system, our scheme would be second order.
\subsection{The fourth order scheme}
In the 3D test calculations it turned out that to obtain the
accuracy which we regard as necessary would require more computer
resources than available to us.
We therefore, and due to the strong recommendations by H.-O.~Kreiss,
tried the ``method of line'' to build higher order schemes.
Since the first results on 3D calculations on the Maxwell equations
and the SU(2)-Yang-Mills equations confirmed the statement, that the
required number of gridpoints per coordinate directions is reduced by
a factor of five~\cite{Kr98pC}, we combined the method of line with
the conformal approach.
\\
In the method of line we formally write 
\begin{equation}
\label{FormaleGleichung}
  \partial_t \underline{f} = 
    \underline{B}(\underline{f},\partial_i\underline{f}),
\end{equation}
where $\underline{B}(\underline{f},\partial_i\underline{f}) = -
\underline{\underline{A}}^i(f) \partial_i \underline{f} +
\underline{b}(f)$, and integrate the ``ordinary differential
equation''~(\ref{FormaleGleichung}) by a scheme to integrate ordinary
differential equations, in our case a fourth order Runge-Kutta scheme:
\begin{eqnarray}
\label{RK4}
  \underline{f}^{l+1}_{i,j,k} & = &
  \underline{f}^{l}_{i,j,k}
  + \frac{1}{6} 
    \left( \underline{k}^l_{i,j,k} 
           + 2 \underline{k}^{l+1/4}_{i,j,k} 
           + 2 \underline{k}^{l+1/2}_{i,j,k} 
           + \underline{k}^{l+3/4}_{i,j,k}   \right),
\end{eqnarray}
where 
\begin{eqnarray*}
  \underline{k}^l_{i,j,k} & = &
  \Delta t \>
  \underline{B}(\underline{f}^{l}_{i,j,k},\partial_i\underline{f}^{l}_{i,j,k})
\\
  \underline{k}^{l+1/4}_{i,j,k} & = &
  \Delta t \>
  \underline{B}(\underline{f}^{l+1/4}_{i,j,k},
                \partial_i\underline{f}^{l+1/4}_{i,j,k}), \quad
  \underline{f}^{l+1/4}_{i,j,k} =
    \underline{f^{l}_{i,j,k}} + \frac{1}{2} \, \underline{k}^l_{i,j,k}
\\
  \underline{k}^{l+1/2}_{i,j,k} & = &
  \Delta t \>
  \underline{B}(\underline{f}^{l+1/2}_{i,j,k},
                \partial_i\underline{f}^{l+1/2}_{i,j,k}), \quad
  \underline{f}^{l+1/2}_{i,j,k} =
    \underline{f^{l}_{i,j,k}} 
    + \frac{1}{2} \, \underline{k}^{l+1/4}_{i,j,k}
\\
  \underline{k}^{l+3/4}_{i,j,k} & = &
  \Delta t \>
  \underline{B}(\underline{f}^{l+3/4}_{i,j,k},
                \partial_i\underline{f}^{l+3/4}_{i,j,k}), \quad
  \underline{f}^{l+3/4}_{i,j,k} =
    \underline{f^{l}_{i,j,k}} + \underline{k}^{l+1/2}_{i,j,k}
\end{eqnarray*}
\\
To carry over the convergence order of the time integration we must
calculate the space derivatives in the source term with appropriate
stencils. 
Best results are obtained, especially if space gradients dominate the
error, by pseudo-spectral methods~\cite{Fo96AP}, which are of infinite 
order in space.
Nevertheless, we do not want to use pseudo-spectral methods, because
we want to keep the freedom to continue the calculation after singular
values appeared at some gridpoints on a slice.
\\
Instead of using spectral decompositions to calculate the space
derivatives, we approximate derivatives by the symmetric fourth order
stencil~(\ref{AblDiskrIV}), which stretches out two gridpoints to each
side.
To ensure stability we can, and actually we have to, add dissipative
terms of higher order.
For systems with constant coefficients theorem~6.7.1 and theorem~6.7.2
of~\cite{GuKA95TD} show stability.
It is also discussed there how to extend results for systems with
constant coefficients to systems with variable coefficients. 
\\
The dissipation term suggested by the theorems is
\begin{eqnarray}
\label{DiskrDiss}
  \lefteqn{
    \partial_x{}^6 \underline{f}^l_{i,j,k} = 
      \frac{1}{(\Delta x)^6}
      \left( \underline{f}^l_{i-3,j,k} - 6 \underline{f}^l_{i-2,j,k}
             + 15 \underline{f}^l_{i-1,j,k} \right. }\hspace{8em}
 \nonumber\\
  & & \left. {}
             - 20 \underline{f}^l_{i,j,k}
             + 15 \underline{f}^l_{i+1,j,k} 
             - 6 \underline{f}^l_{i+2,j,k} + \underline{f}^l_{i+3,j,k}
    \right),         
\end{eqnarray}
which has a seven point stencil stretching out three gridpoints on each side.
By adding $\sigma Q_2 := \frac{\sigma }{64 \, N} (\Delta x)^5 \sum_{i=1}^{N}
\partial_i{}^6 f$ with a sufficiently large $\sigma$ to each
evaluation of $\underline{B}(\underline{f},\partial_i\underline{f})$
we got a stable scheme.
Numerical experiments showed that large values for $\sigma$ require
small time steps $\Delta t$ for stability, therefore $\sigma$ should be
chosen large enough, but as small as possible.
The test cases have been run with $\sigma=2$, other data may
require larger values.
\\
We also looked at a scheme which adds dissipation from a five point
stencil after each full fourth order Runke-Kutta step, namely the term 
$\sigma \bar Q_2 = \frac{- \sigma}{16 \, N} (\Delta x)^4
\sum_{i=1}^{N} \partial_i{}^4 f$ with
\begin{equation}
\label{DiskrDissIV}
  \partial_x{}^4 \underline{f}^l_{i,j,k} = 
    \frac{1}{(\Delta x)^4}
    \left( \underline{f}^l_{i-2,j,k}
           - 4 \underline{f}^l_{i-1,j,k} 
           + 6 \underline{f}^l_{i,j,k}
           - 4 \underline{f}^l_{i+1,j,k} 
           +   \underline{f}^l_{i+2,j,k} 
    \right).
\end{equation}
For $\sigma=2$ this also seems to be stable but it only yields a
third order scheme and was therefore not pursued any farther.
\\
In the numerical implementation we use four grids to store the
variables, the gauge source functions, and the intermediate values. 
And again, even if we calculated the gauge source functions by
some global fourth order procedure, the scheme would stay fourth
order.
\subsection{The boundary treatment}
\label{Randbehandlung}
There are two kind of boundaries to be dealt with.
The first kind, which we call false boundaries, comes from reducing
the grid size, but not the grid dimension, by assuming discrete
symmetries.
In our case these are the periodic boundaries in the $y$ and $z$
directions in asymptotically A3 spacetimes and the boundaries
coinciding with the symmetry planes in octant\footnote{Octant mode is
  obtained when assuming a mirroring symmetry with respect to the x,
  y, and z plane through the origin.} mode.
On the false boundaries $\Omega$ may assume any value.
The second kind of boundaries, which we call true boundaries, restrict
the treated range of the conformal spacetime.
To avoid any significant influence of their treatment onto the physical
spacetime they must be placed into regions with $\Omega<0$.
\\
In the first part of this series we have described how we change the
equations near the true boundaries.
For the tests we used the modification~(I/19), which freezes the time
evolution near the true boundaries.
\\
Before we take a time step, we extend the grid on the true boundaries
by the stencil width by a first order extrapolation.
At false boundaries we extend by copying the values from the
corresponding gridpoints in the interior.
Then we take the time step and, of course, loose the just created
gridpoints again.
\subsection{Controlling the time step}
It is well-known that the Courant-Friedrichs-Lewy condition, which
states that the numerical domain of dependence must contain the
analytic domain of dependence, is necessary for stability of symmetric
hyperbolic schemes.
This requirement restricts the maximal size of the time step.
To calculate a linear approximation to its size we take the forward
light cone at all gridpoints $(i,j,k)$ and calculate the time $\Delta
t_{\mathrm{min}\>i,j,k}$ of its earliest intersection with the
boundaries given by the neighbouring gridpoints with coordinate values
$x_{i-1}$, $x_{i+1}$, $y_{j-1}$, $y_{j+1}$, $z_{k-1}$, and $z_{k-1}$.
The minimum over the grid $\Delta t_{\mathrm{min}} := min_{(i,j,k)} \Delta
t_{\mathrm{min}\>i,j,k}$ would be the maximally allowed time step for
the pure rotated Richtmyer scheme.
Stability requirements not caused by the CFL-condition and the use of
wider stencils modify the allowed time step to a multiply $q$ of
$\Delta t_{\mathrm{min}}$. 
We ran all calculations reported in this paper with the value $q=1/2$.
\\
The time step control by linearly approximating the light cone has
consequences for what we can expect, when we check convergence of a
scheme by subtracting the results from runs with different coarseness 
of the grids at corresponding time steps.
Since the size of the time steps deviates by second order,
corresponding slices have time coordinates which deviate by second
order.
The sequence of the differences will therefore seem only second order
convergent, even if the scheme really converges with higher order.
%
%

%
%
\section{The tests performed}
\label{Tests}
\subsection{The exact solutions used for testing}
\label{ExactLsg}
To test the code we reproduce exact solutions of the conformal field 
equations by prescribing the data and the gauge source functions from
the exact solutions.
To better understand the test runs we shortly recall those properties
of the used exact solutions which are important for us.
\\[0.3em]
The first class of exact solutions, which we use in the 2D and 3D test
runs, are the so-called asymptotically A3 solutions~\cite{Hu98ma}.
They are given by
\begin{equation}
\label{A3artigeLsg}
\left( \begin{array}{cc}
         g_{ab} 
       \\ 
         \Omega 
       \end{array}
\right) = 
\left( \begin{array}{cc}
         4 \, \sqrt{\frac{2}{t_s{}^2+x_s{}^2}} \, e^M 
           \left( -dt_s{}^2 + dx_s{}^2 \right)
         + \frac{1}{2} \left( t_s{}^2 + x_s{}^2 \right)
           \left( e^W dy_s{}^2 + e^{-W} dz_s{}^2 \right) 
       \\[0.5em]
         \frac{1}{4} \left( t_s{}^2 - x_s{}^2 \right)
       \end{array}
\right),
\end{equation}
where $M$ and $W$ are certain functions of $t_s$ and $x_s$.
\\
If we set
\begin{equation}
\label{A3Lsg}
  (M,W)=(0,0),
\end{equation}
we obtain the A3 solution.
\\
The choice
\begin{equation}
\label{WieA3Lsg}
  (M,W) = 
  \left( -\frac{\left(t_s{}^2+x_s{}^2\right)^2}{256},
         \frac{t_s{}^2-x_s{}^2}{8} 
  \right)
\end{equation}
yields a solution, which contains gravitational
radiation~\cite{Fo97sa}.
Figure~\ref{KonfRZA3artig} shows a conformal diagram of asymptotically 
A3 solutions with the periodic $y$ and $z$ coordinates suppressed.
\begin{figure}[htbp]
  \begin{center}
      \begin{minipage}[t]{6.0cm}
        \input A3KonfRZ.pstex_t
        \vskip0.5em
        \caption{\label{KonfRZA3artig}Conformal spacetime for the
          asymptotically A3 spacetimes with y and z coordinate suppressed}
      \end{minipage}
    \hspace{4em}
      \begin{minipage}[t]{7cm}
        \input KonfMink.pstex_t
        \vskip0.5em
        \caption{\label{KonfMink}Conformal Minkowski spacetime with z
          coordinate suppressed}
      \end{minipage}
  \end{center}
\end{figure}
We start at $t_0=-1$ and integrate until $t_1=-1/2$. 
As long as $t_0$ and $t_1$ are negative, their choice is completely
arbritrary.
The origin $(0,0)$ is singular, the components of the metric and the
curvature become singular there.
In principle we could continue the calculation beyond $t_1$ towards
$(t_s,x_s)=(0,0)$, and for test purposes we have done so.
The CFL condition then forces us to use smaller and smaller time steps 
and we approach, but never reach, the origin.
\\
In the calculations reported here we have hidden the symmetries in the 
2D calculations by the transformation 
\begin{equation}
\left( \begin{array}{ccc}
         t_s \\ x_s \\ y_s 
       \end{array}
\right) = 
\left( \begin{array}{ccc}
         t
       \\ 
         x \left\{ 1 + 
                   \frac{1}{2} \left( x - 1 \right)
                   \left[ \sin\left( \pi
                                     \frac{y-y_{\rm min}}
                                          {y_{\rm max}-y_{\rm min}} 
                                      
                              \right)
                  \right]^2
           \right\}
       \\
         y
       \end{array}
\right)
\end{equation}
and in the 3D calculations by the transformation
\begin{equation}
\left( \begin{array}{cccc}
         t_s \\ x_s \\ y_s \\ z_s
       \end{array}
\right) = 
\left( \begin{array}{cccc}
         t
       \\ 
         x \left\{ 1 + 
                   \frac{1}{2} \left( x - 1 \right)
                   \left[ \sin\left( \pi
                                     \frac{y-y_{\rm min}}
                                          {y_{\rm max}-y_{\rm min}} 
                                      
                              \right)
                  \right]^2
                  \left[ \sin\left( \pi
                                     \frac{z-z_{\rm min}}
                                          {z_{\rm max}-z_{\rm min}} 
                                      
                              \right)
                  \right]^2
           \right\}
       \\
         y
       \\
         z
       \end{array}
\right).
\end{equation}
\\[0.3em]
Another exact solution, we have tested against, is the conformal
representation of Minkowski spacetime, which is in spherical
coordinates $(t,r,\theta,\varphi)$ given by
\begin{equation}
\left( \begin{array}{cc}
         g_{ab} 
       \\ 
         \Omega 
       \end{array}
\right) = 
\left( \begin{array}{cc}
         - dt^2 + dr^2 
         + (\sin r)^2 \left[ d\theta^2 + (\sin \theta)^2 d\varphi^2 \right] 
       \\[0.5em]
         2 \left/
         \sqrt{ 
           \left( 1 + \left( \tan \frac{t+r}{2} \right)^2 \right)
           \left( 1 + \left( \tan \frac{t-r}{2} \right)^2 \right)
          } \right.
       \end{array}
\right).
\end{equation}
Of course, in the actual calculation we have transformed to Cartesian
coordinates, where $g_{ab}$ is no longer diagonal and the spherical
symmetry is hidden.
It should also be mentioned that this conformal solution is time
dependent, since the conformal factor $\Omega$ is.
But the time dependence is in a certain sense very weak, already the
test calculations for the spherically symmetric case in~\cite{Hu93nu}
have shown, that the accuracy obtained is order of magnitudes better
than what is obtained in interesting cases.
This statement is confirmed by the results of
subsection~\ref{Tests3D}.
\\
Due to our boundary treatment we can expect the reproduction of the
exact solution and the propagation of the constraints only at
gridpoints representing physical spacetime and \Scri{}, although the
solutions given exist outside the physical region.
Therefore, we apply the measures for the quality of a solution, as
defined in subsection~\ref{QuaMass}, only at the gridpoints
representing physical spacetime.
In the used representation of conformal Minkowski spacetime we reach
timelike infinity $i^+$ after a finite conformal time.
Hence, the number of gridpoints representing physical spacetime
decreases to zero.
To have a significant part of the grid available to evaluate our
accuracy measures we compare on the slice half way to $i^+$, although
the calculations have been continued beyond $i^+$.
Even when surpassing $i^+$ there is no change in the convergence of
the scheme.
By being able to cover timelike infinity with gridpoints we have a
powerful numerical tool to study the fall-off behaviour of radiative
quantities for very long times~\cite{Hu96na}.
\\[0.3em]
Another solution which is commonly used for testing codes in numerical 
relativity is the Schwarzschild solution.
So, why is this code not yet tested against the Schwarzschild solution?
The reason is simple: 
The author is not aware of any explicit solutions of the
conformal field equations representing the Schwarzschild-Kruskal
spacetime, which is regular at both ${\cal S}$s and which is smoothly
extendible across both ${\cal S}$s.
By Birkhoff's theorem, the theorems in~\cite{AnCA92ot}, and the
theorems in~\cite{Fr91ot} we know that all the requirements above can
be achieved, except that the solution is not explicit.
When describing the initial data solver part of the code, we are going
to prescribe spherically symmetric free functions with two
${\cal S}$s, which then necessarily yield data for the
Schwarzschild-Kruskal spacetime.
\subsection{Measures for the quality of a numerical solution}
\label{QuaMass}
Without doubt, the best measure for the quality of a numerical
solution is the difference to the exact solution:
\begin{equation}
  \underline{\Delta}_{i,j,k} := 
    \underline{f}_{i,j,k} - \underline{f}_{\mathrm{exact}\>i,j,k},
\end{equation}
where $\{i,j,k\}$ denotes gridpoints representing physical spacetime or 
null infinity.
Since gravitational radiation is given by the Newman-Penrose quantity
$\psi_4$ of the conformal Weyl tensor $d_{abc}{}^d$ evaluated at
\Scri{}, which is essentially a polynomial expression in $E_{ab}$ and
$B_{ab}$, the error in $E_{ab}$ and $B_{ab}$ is also a measure for the 
error in the gravitational radiation.
As we do not prescribe all variables when providing an exact solution, 
but numerically calculate $\RII_{ab}$, $E_{ab}$, and $B_{ab}$ by
solving elliptic equations, the comparison of all variables with the
exact solution is very time consuming, especially since the AMG
library does not run in parallel.
We therefore compare with the exact solution only once towards the end
of a calculation.
On the intermediate time steps we use what we call pseudo-difference,
that is we calculate $( h_{ab}$, $k_{ab}$, $\gamma^a{}_{bc}$,
$\ROI_a$, $f_{\RII\,ab}$, $\Omega f_{E\,ab}$, $f_{B\,ab}$, $\Omega$,
$\Omega_0$, $\Omega_a$, $\omega )$ from the exact solution given by
$(g_{ab}, \Omega)$ and compare with $(h_{ab}$, $k_{ab}$,
$\gamma^a{}_{bc}$, $\ROI_a$, $\Omega \RII_{ab}$, $\Omega^2 E_{ab}$,
$\Omega B_{ab}$, $\Omega$, $\Omega_0$, $\Omega_a$, $\omega )$.
Comparisons of the measure ``difference'' and ``pseudo-difference''
have been performed and the comparisons showed that the measures are
measures exchangeable with respect to the relative error, which we
define as
\begin{equation}
  \underline{\Delta}_{\mathrm{rel}\>i,j,k} := 
    \frac{\underline{f}_{i,j,k} -
      \underline{f}_{\mathrm{exact}\>i,j,k}}
      {\max_{(i,j,k)}(1,|\underline{f}_{i,j,k}|)}.
\end{equation}
With respect to the absolute error the results differ, since e.~g.\
$\Omega^2 E_{ab}$, which is used in pseudo-difference, may be
significantly smaller than $E_{ab}$, and since the variables
representing second derivatives of the metric, like $E_{ab}$, tend to
dominate the absolute error.
\\[0.3em]
We also monitor the violation of the constraints.
In later runs, which do not reproduce an exact solution, the
convergence of the violation of the constraints to zero is our main
criteria for consistency of the numerical time evolution with the
Einstein equation, although we cannot conclude from the size of the
numerical constraint violation to the quality of the numerical
solution~\cite{Ch91co}.
\\[0.3em]
When debugging and analysing the performance of the code we look at
the measures pointwise by looking at surface plots on well-chosen
two-dimensional slices.
Although this implies looking at hundreds of plots to get
representative samples, we regard it as unavoidable for a thorough
testing and tuning of the code.
Since we cannot present hundreds of figures in a paper we must
condense for the presentation in this paper.
To do so, we plot
\begin{equation}
\label{NormVerletzung}
  \|\Delta\|(x) := 
    \frac{1}{M} 
      \sqrt{\sum_{l=1}^{M} 
        \int_{\{y,z\}} \Big(\Delta_l(x,y,z)\Big)^2 dy \, dz}
  \sim
    \frac{1}{M} 
      \sqrt{\frac{1}{N_y N_z} 
        \sum_{l=1}^{M} \sum_{\{j,k\}} \Big(\Delta_{l\>i,j,k}\Big)^2 },
\end{equation}
where $M$ is the number of the entries in the vector
$\underline{\Delta}=\Delta_l$ of quantities which should go to zero.
\subsection{Tests of the 2D code}
\begin{figure}[htbp]
  \begin{center}
      \begin{minipage}[t]{7cm}
        \epsfxsize=7cm
        \centerline{\epsfbox{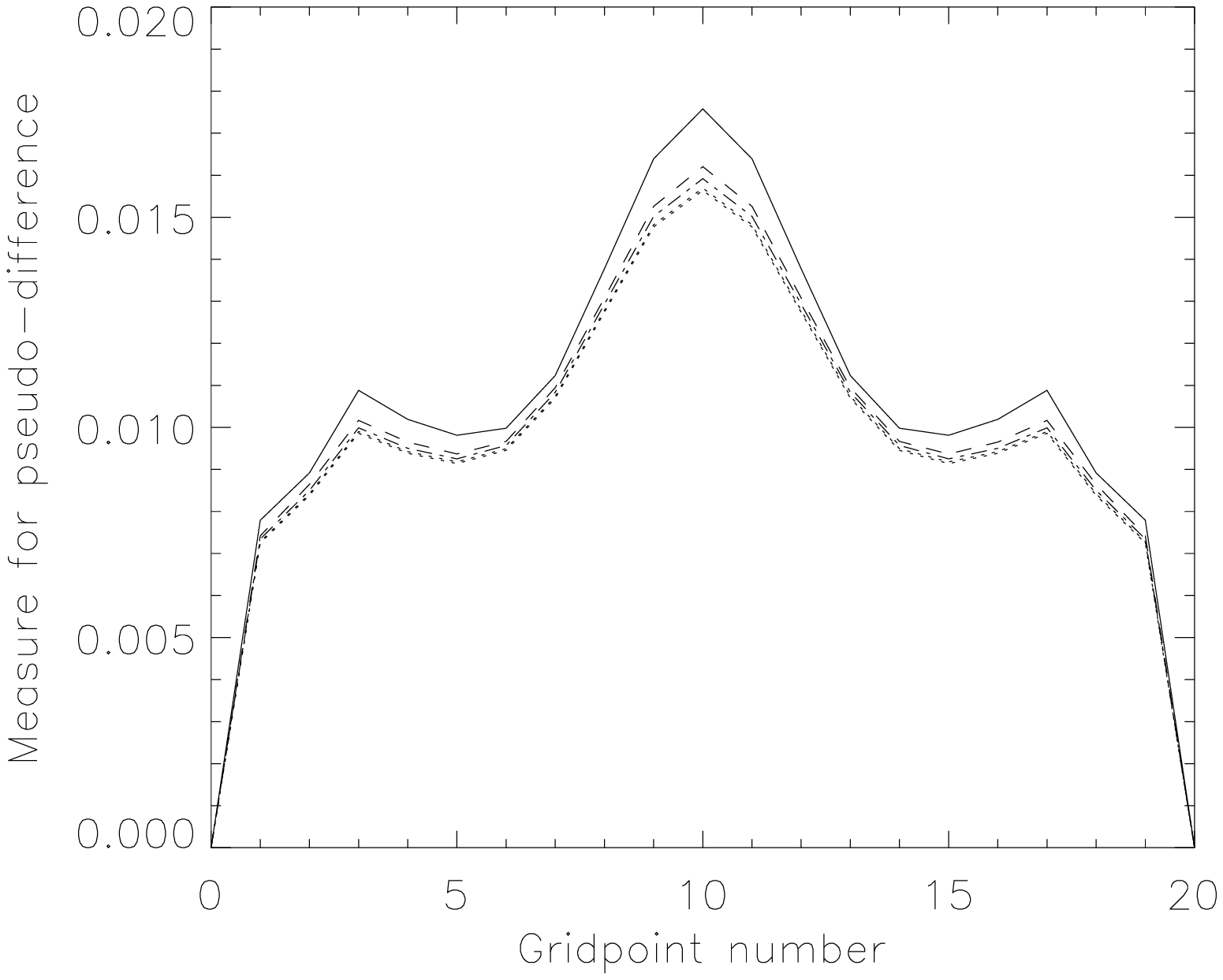}}
        \caption{\label{IIteA3KonvDelta}Convergence against an A3 like
          solution for the 2nd order scheme in 2D}
      \end{minipage}
    \hspace{2em}
      \begin{minipage}[t]{7cm}
        \epsfxsize=7cm
        \centerline{\epsfbox{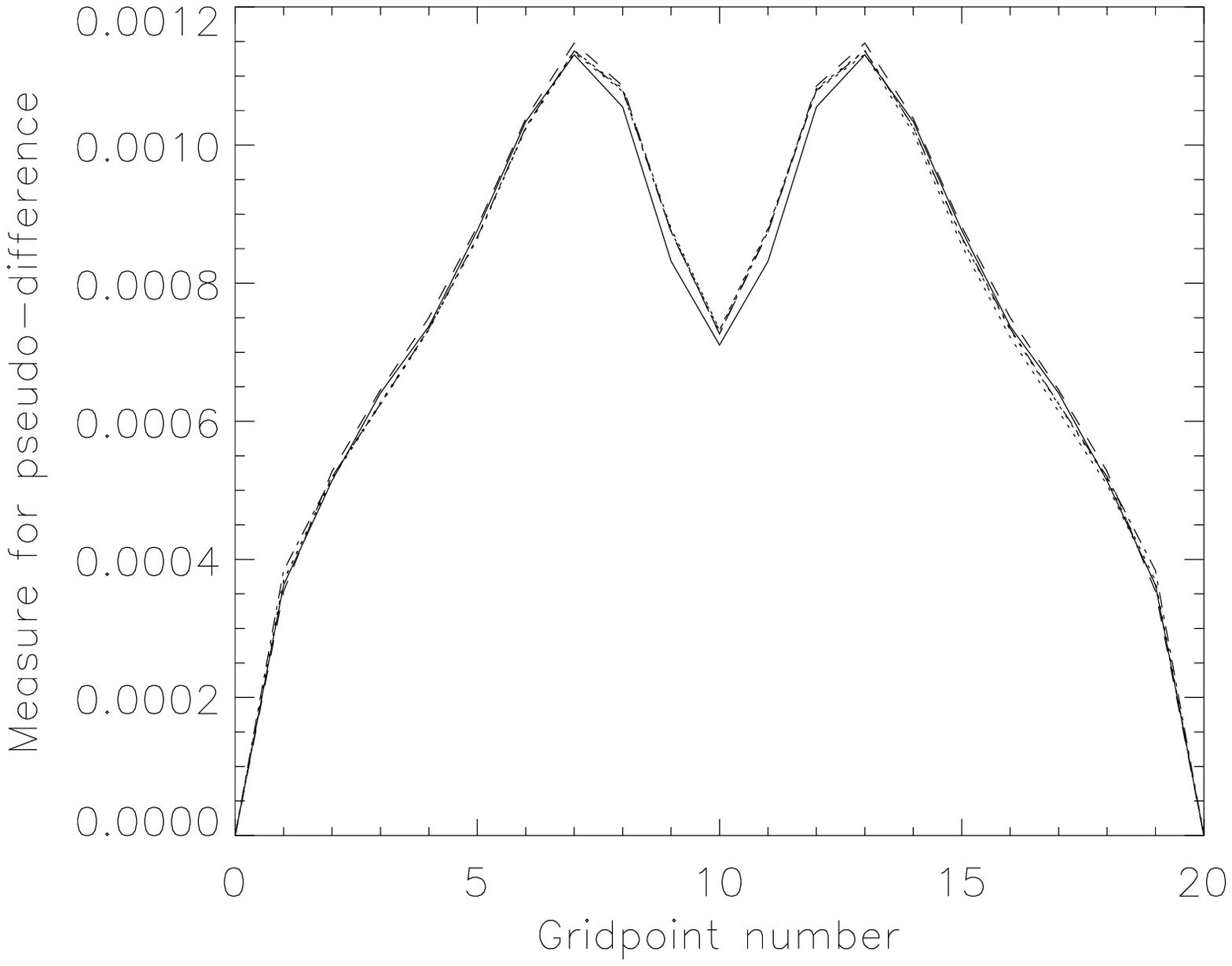}}
        \caption{\label{IVteA3KonvDelta}Convergence against an A3 like
          solution for the 4th order scheme in 2D}
      \end{minipage}
  \end{center}
\end{figure}

\begin{figure}[htbp]
  \begin{center}
      \begin{minipage}[t]{7.0cm}
        \epsfxsize=7cm
        \centerline{\epsfbox{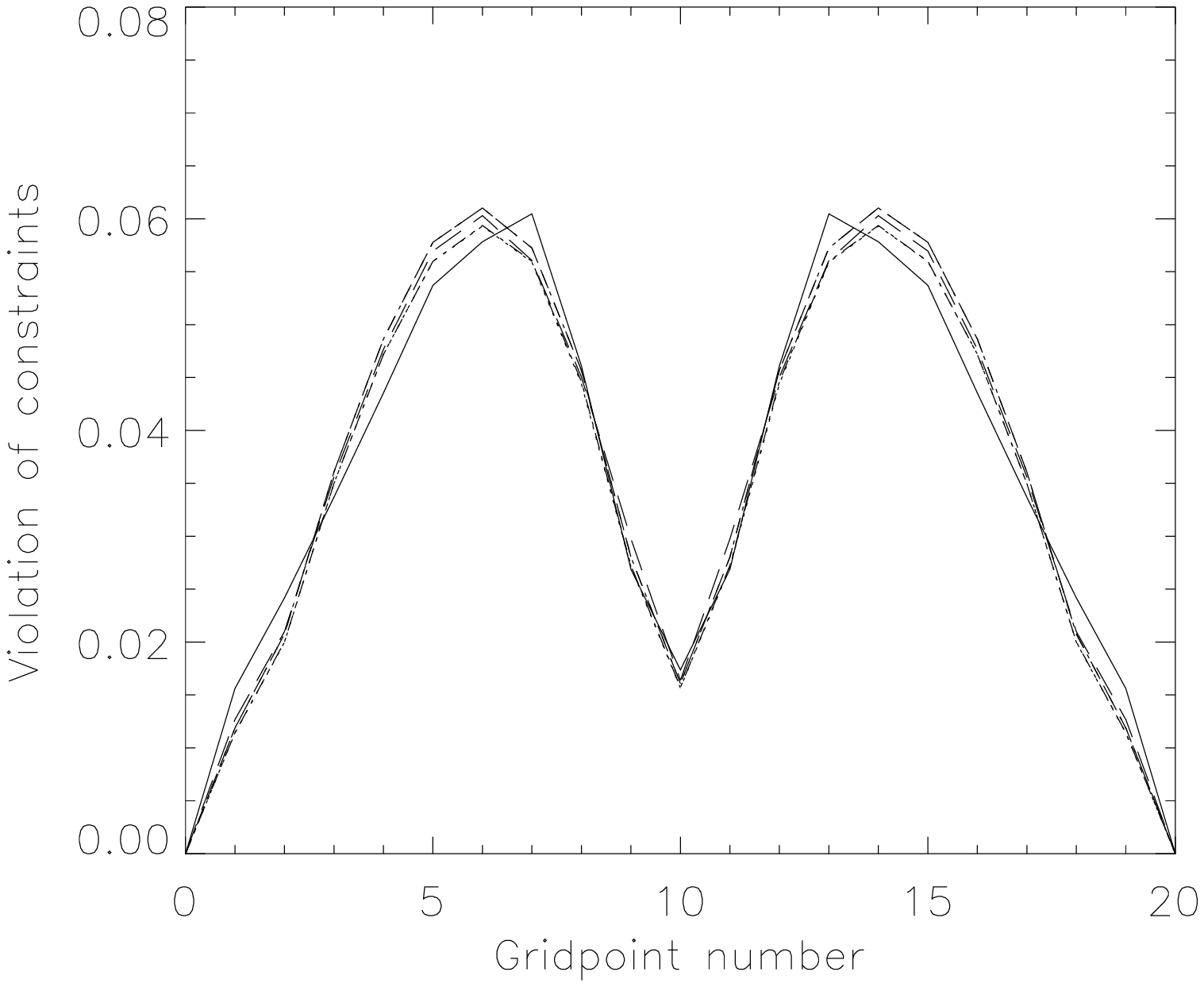}}
        \caption{\label{IIteA3KonvConstr}Convergence of the
          violation of the constraints in an A3 like solution for the 2nd
          order scheme in 2D}
      \end{minipage}
    \hspace{2em}
      \begin{minipage}[t]{7cm}
        \epsfxsize=7cm
        \centerline{\epsfbox{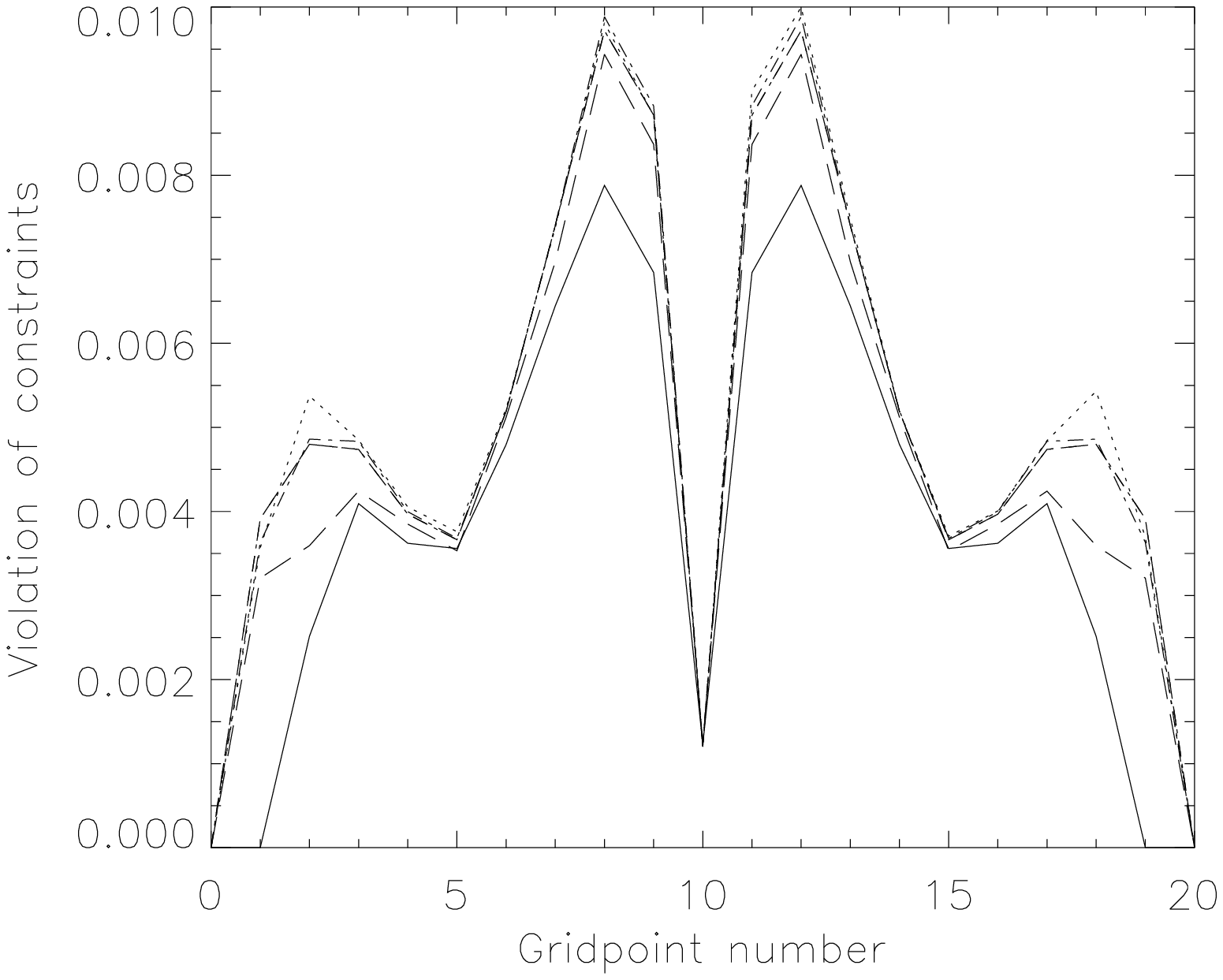}}
        \caption{\label{IVteA3KonvConstr}Convergence of the
          violation of the constraints in an A3 like solution for the 4th
          order scheme in 2D}
      \end{minipage}
  \end{center}
\end{figure}

In the first tests we ran we did not hide the obvious Killing vector
$\partial_{y_s}$ in expression~(\ref{A3artigeLsg}). 
The only remarkable thing to report about is, that in this case we
could find a stable treatment of the true boundaries without changing
the equations near the true boundaries as described in
subsection~\ref{Randbehandlung}.
For solutions with hidden Killing vectors the first became unstable,
forcing us to introduce the later treatment.
\\
From the numerical viewpoint, solution~(\ref{A3Lsg}), the A3 solution
without gravitational radiation, and solution~(\ref{WieA3Lsg}), a
solution with gravitational radiation, behave very similar.
\\[0.3em]
Figure~\ref{IIteA3KonvDelta} shows the convergence of the second order 
scheme in the measure~(\ref{NormVerletzung}), where $\Delta_l$ denotes 
the pseudo-difference to the exact solution at $t_1=-1/2$.
The lines plotted correspond to calculations with $40^2$ (solid),
$80^2$ (dashed), $160^2$ (dotted), $320^2$ (dotted), and $640^2$
gridpoints (dotted).
The values are scaled in such a way that the lines would coincide, if
the convergence were exactly second order.
The gridpoint numbers are with respect to the output grid which has a
constant number of gridpoints independent of the grid on which the
calculation is performed.
Obviously in the calculation with $40^2$ gridpoints higher order terms 
still make a significant contribution to the error, the solid line
does not coincide with the dashed line.
For finer grids we get the expected rate of convergence.
\\
Since the size of the error measure is not immediately related to the
maximum norm of the error, we give it as well:
The maximum of the absolute error drops in good agreement with
the convergence rate from $4.44$ in the $40^2$ run to $0.01$ in the
$640^2$ run.
The variable with the largest error is in both cases $E_{11}$.
The variable with the largest relative error is $k_{11}$, its value
drops, again in good agreement with the convergence rate,  from 17\%
to 0.07\%.
\\
Figure~\ref{IIteA3KonvConstr} shows the scaled measures for the
violation of the constraints.
Since all the curves almost coincide, the violation of the constraints 
converges to zero in excellent agreement with the convergence order.
\\[0.3em]
Figure~\ref{IVteA3KonvDelta} is the analogue to
figure~\ref{IIteA3KonvDelta} for the fourth order scheme, but with a
fourth order scaling.
We observe that already for the $40^2$ run the error is dominated by
fourth order contributions.
Already for this coarsest grid, the error measure is one order of
magnitude smaller than what we obtained for the second order scheme.
This result is also confirmed by the figures for the maximum norm,
which are:
The maximal absolute error drops from $1.35$ to $2.1\times10^{-5}$ and 
is found as in the second order scheme in the variable $E_{11}$.
In contrast to the second order scheme, the maximal relative error
appears for the variable $E_{12}$ and decreases from $6.2$\% to
$1.0\times10^{-4}$\%.
This is all in excellent agreement with fourth order convergence.
\\
Good, but not excellent agreement with fourth order convergence can
be found in figure~\ref{IVteA3KonvConstr}, which shows the convergence
of the violation of the constraints.
The scaled curves for the $40^2$ (solid line) and the $80^2$ (dashed
line) deviate from the curves for runs with finer grids ($160^2$ and
$320^2$), which coincide very well.
The dashed-dotted line representing the $640^2$ run deviates slightly
around gridpoint $2$ and $18$ from the $160^2$ and the $320^2$ runs.
This is not serious, in this run we have reached the lower accuracy
bound at which rounding errors inherited from the ``$\Omega$ divider'' 
become significant (see also the discussion at the end of
subsection~\ref{DivdurchOm}).
\subsection{For 3D}
\label{Tests3D}
\subsubsection{Minkowski}
\begin{figure}[htbp]
  \begin{center}
      \begin{minipage}[t]{7cm}
        \epsfxsize=7cm
        \centerline{\epsfbox{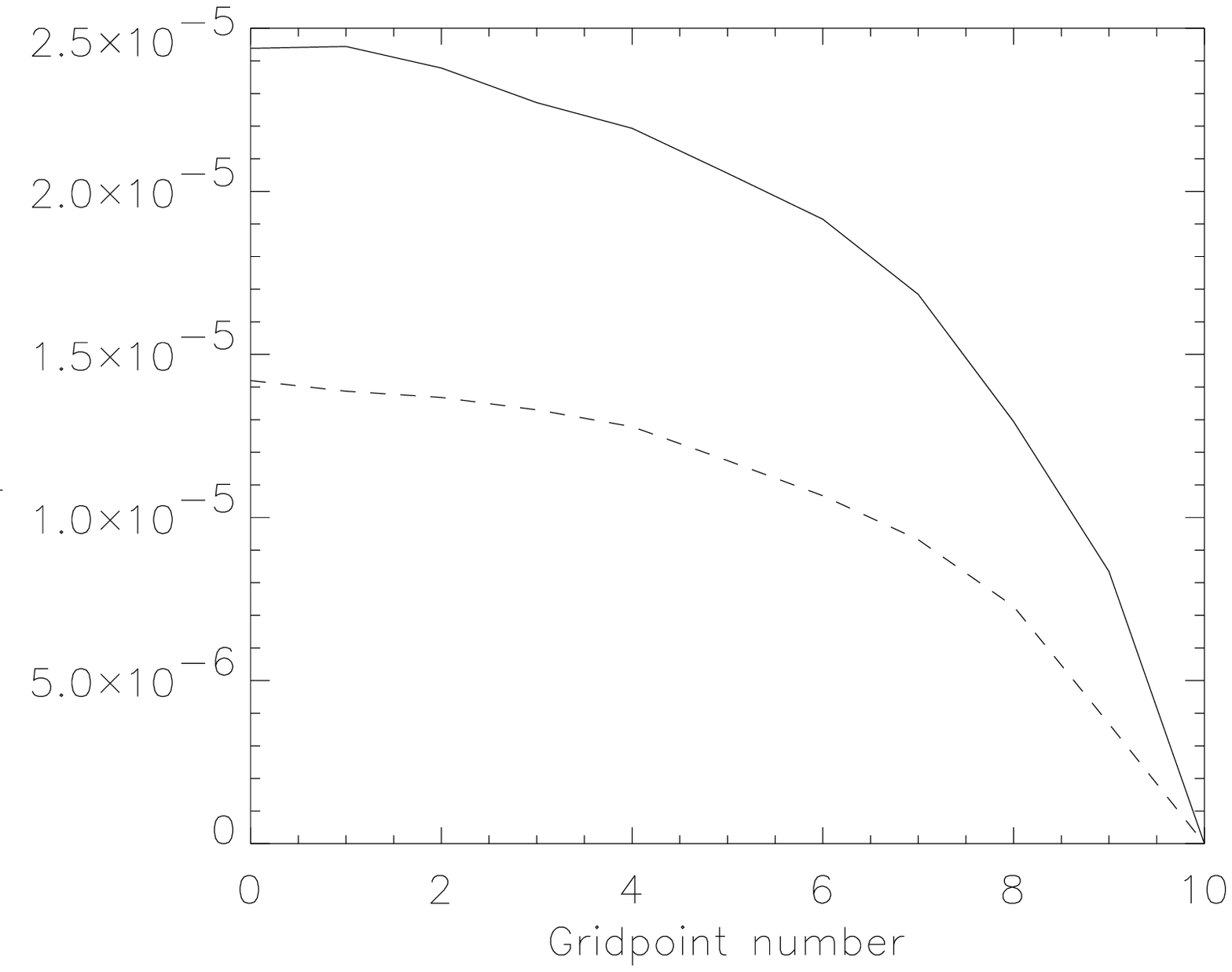}}
        \caption{\label{IIteMinkKonvDelta}Convergence against
          the Min\-kow\-ski solution for the 2nd order scheme in 3D}
      \end{minipage}
    \hspace{2em}
      \begin{minipage}[t]{7cm}
        \epsfxsize=7cm
        \centerline{\epsfbox{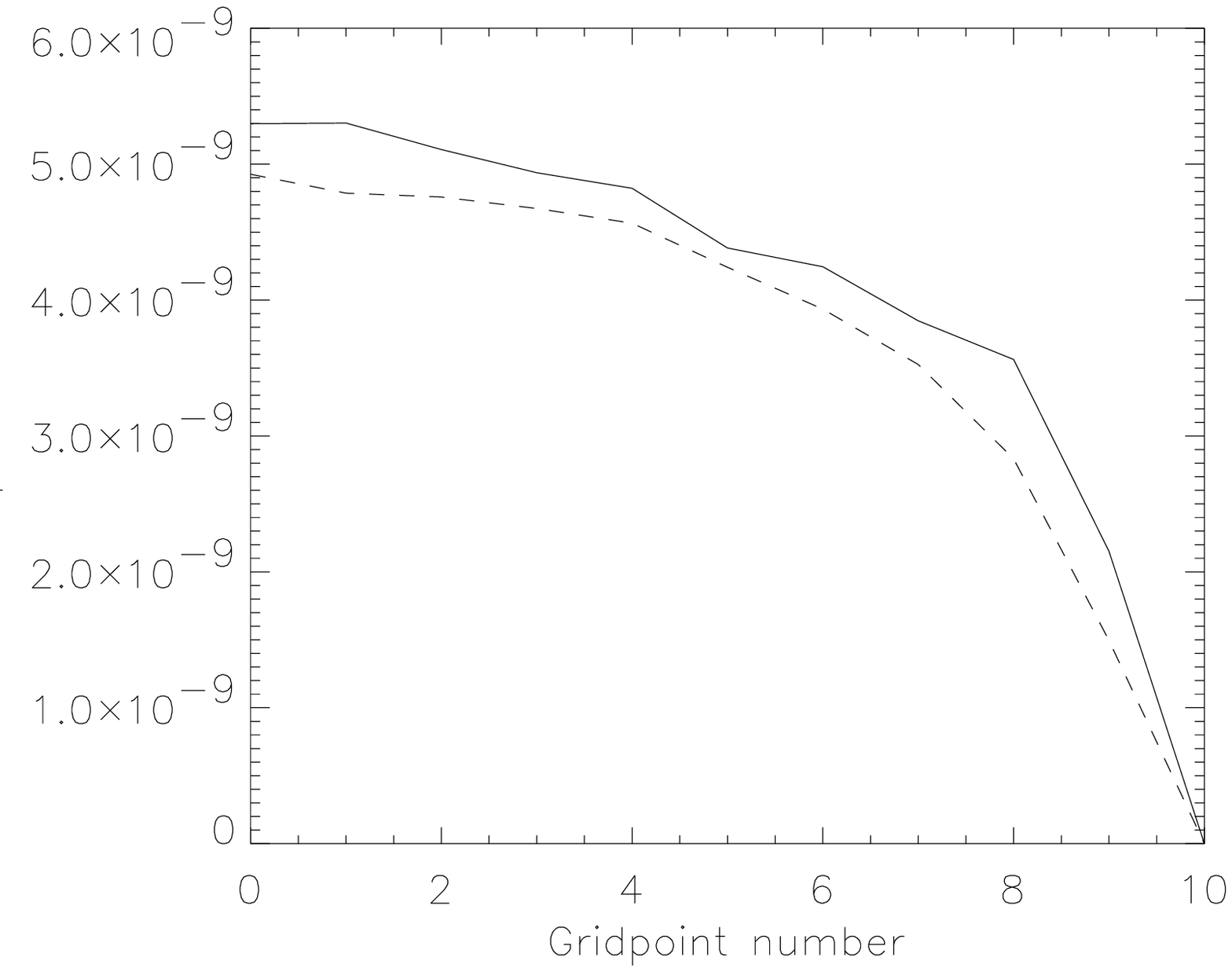}}
        \caption{\label{IVteMinkKonvDelta}Convergence against
          the Min\-kow\-ski solution for the 4th order scheme in 3D}
      \end{minipage}
  \end{center}
\end{figure}

Figure~\ref{IIteMinkKonvDelta} and figure~\ref{IVteMinkKonvDelta} show 
the measure for the pseudo-difference for a second and a fourth order
3D run with $50^3$ (solid line) and $100^3$ (dashed line) with data
for the Minkowski spacetime.
The solid line is scaled such that with exact convergence it would
lie on the dashed line.
Although the runs have been done on a full grid, despite the octant
symmetry, the figures are based on an octant.
We immediately see, that, due to the low number of gridpoints, the
error of both schemes has contributions from higher order terms.
These higher order contributions are more significant in the second
order run.
We do not regard this failure as very serious as the achieved accuracy 
is already extremely high.
\\
In the $100^3$ run the maximum of the absolute error is extremely
small, it is $6.6\times10^{-5}$ for $\RII_{33}$ in the second order
run and $2.2\times10^{-8}$ for $\RII_{33}$ in the fourth order run.
The exactly same numbers hold for what we have defined as the maximal
relative error.
\\
If we calculate under the assumption of second order convergence, what 
grid size we would need to achieve in a second order run the same
accuracy with respect to the maximum of the relative error as in
the $100^3$ fourth order run, we get a hypothetical $5400^3$ grid.
Due to the already very small error in both schemes we regard that
estimate as meaningless for practical purposes.
\subsubsection{A3}
\begin{figure}[htbp]
  \begin{center}
      \begin{minipage}[t]{7.1cm}
        \epsfxsize=7cm
        \centerline{\epsfbox{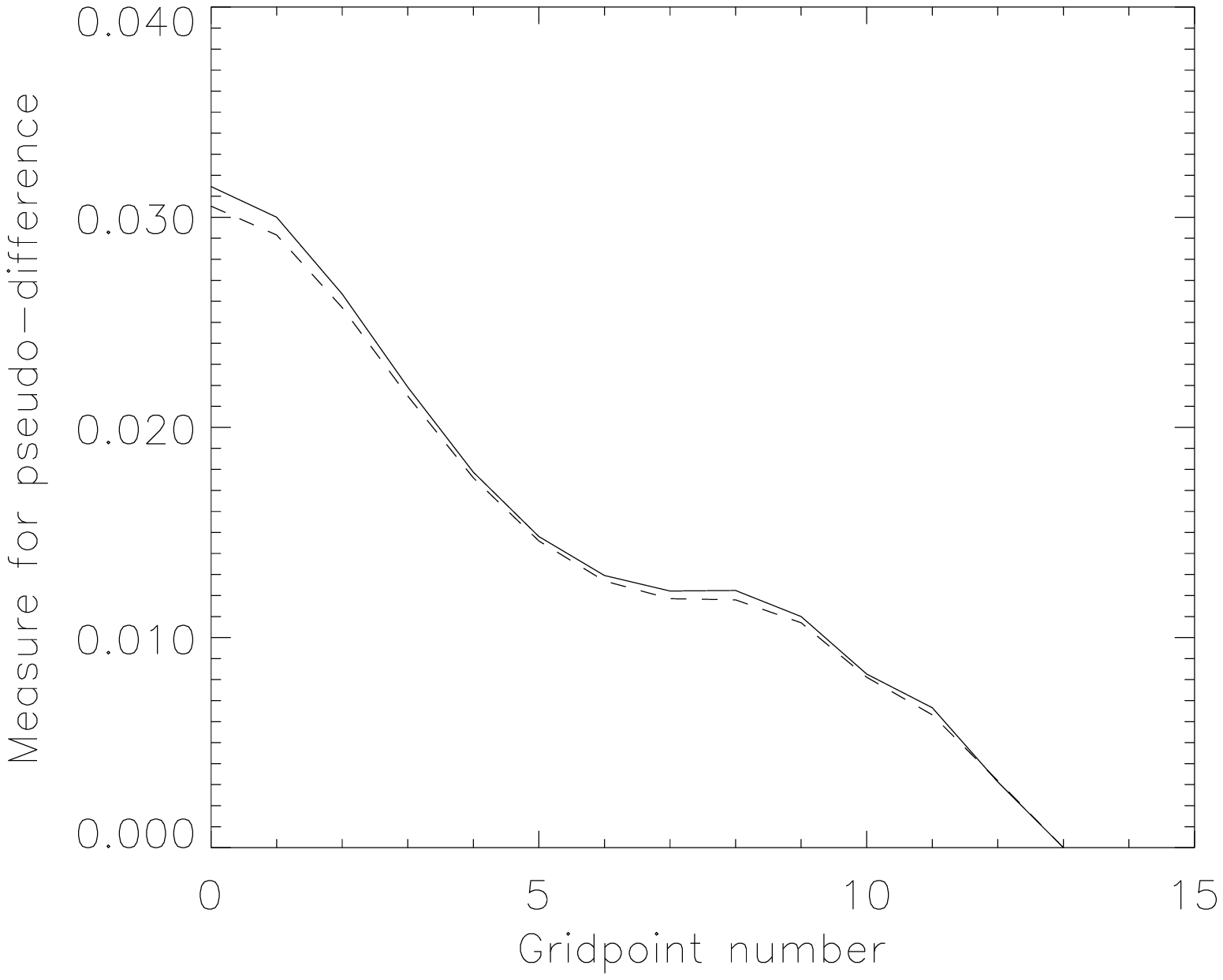}}
        \caption{\label{IIteA33DKonvDelta}Convergence against an A3 like
          solution for the 2nd order scheme in 3D}
      \end{minipage}
    \hspace{2em}
      \begin{minipage}[t]{7.1cm}
        \epsfxsize=7cm
        \centerline{\epsfbox{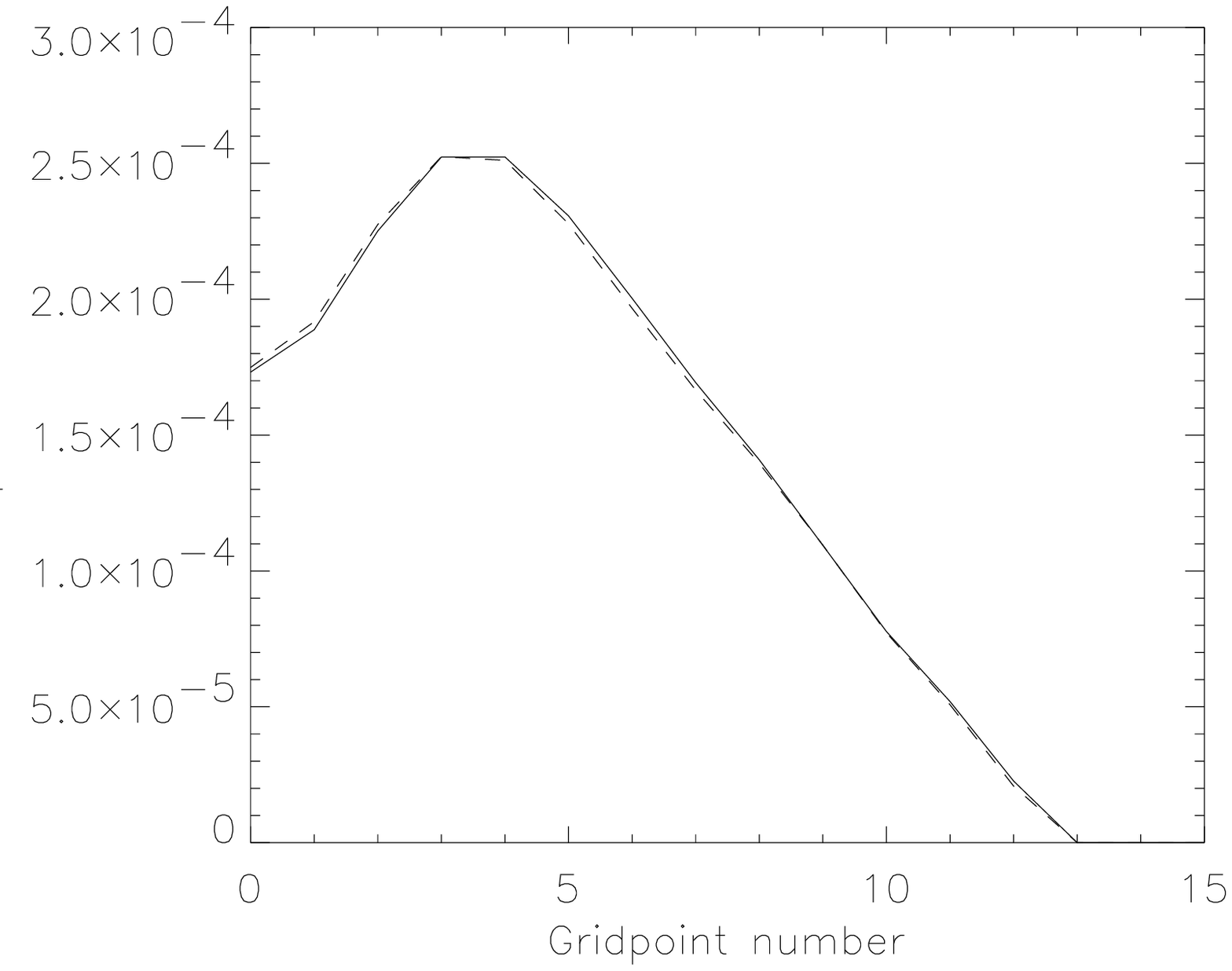}}
        \caption{\label{IVteA33DKonvDelta}Convergence against an A3 like
          solution for the 4th order scheme in 3D}
      \end{minipage}
  \end{center}
\end{figure}

Figure~\ref{IIteA33DKonvDelta} and figure~\ref{IVteA33DKonvDelta} show 
the measure for the pseudo-difference for a second and a fourth order
3D run with $50^3$ (solid line) and $100^3$ (dashed line) on an A3
like spacetime with hidden $y$ and $z$ Killing symmetries.
\\
Again the solid line is scaled such that with exact convergence it would
lie on the dashed line and we only plot an octant.
The order of convergence is in excellent agreement with the
prediction, the contribution of higher order terms to the error is
almost negligible.
\\
With the $100^3$ grid the maximum of the absolute error is $0.41$ for
$E_{11}$ in the second order run and $0.077$ for $E_{11}$ in the
fourth order run.
The largest relative error is $5.8\%$ for $B_{11}$ in the second order
run and $0.24\%$ for $E_{23}$ in the fourth order run.
\\
If we again calculate under the assumption of second order convergence, what 
grid size we would need to achieve in a second order run the same
accuracy with respect to the maximum of the relative error as in
the $100^3$ fourth order run, we get a hypothetical $490^3$ grid.
With respect to the $L_2$-Norm of the pseudo-difference we would even
need $1000^3$ gridpoints.
%
%
%

%
%
\section{Conclusion}
\hskip-\parindent{}%
In this paper we have described a method and its numerical
implementation to derive a complete set of data for the conformal
field equations from a minimal set which works for all scenarios
described in subsection III/B of~\cite{Hu98bh}.
\\
Using data derived from a minimal set given by exact solutions we
have tested our time integration code and compared a second with a fourth
order scheme.
This comparison turned out to be disastrous for the second order
scheme.
To reproduce the accuracy of a maximal relative error smaller than one 
percent in a situation with gravitational radiation, as achieved
by a $100^3$ grid in the fourth order run, which needs $1.7$~GB of
memory and 7~hours on 16~processors of a origin 2000, we would need
more than 200~GB memory and 3000~hours on 16~processors for a run with 
the second order scheme. 
\section*{Acknowledgement}
\hskip-\parindent{}%
I would like to thank B.~Schmidt and J.~Frauendiener for many fruitful
discussions.
In particular I would like to thank H.~Friedrich for his extensive
help and support.
\\
On numerical questions I got indispensable help from H.-O.~Kreiss, who
pointed out to me the superiority of the methods of line, and
K.~St\"uben from the Gesellschaft f\"ur Mathematik und
Datenverarbeitung, who put the Algebraic Multigrid Library at my
disposal and was very helpful whenever I had ``elliptic problems''. 
%
%
%
%

\appendix
\section{Remarks about the computational science aspect of the
  code}
\label{ComScieAsp}
We now give a short review of the computational science aspect of the
code.
\\
To be able to do calculations of spacetimes with low symmetries,
especially 3D calculations, it is highly recommendable
to do the resource intensive parts of the calculations in parallel.
On the other side, development of parallel program code tends to take
much longer than writing serial code.
It is therefore advantageous to be able to execute serial as well as
parallel sections in one program, one then has the freedom to program
less resource intensive tasks serial and to save human resources.
The most used library for parallel computing, MPI, does not allow to
combine parallel and serial code, since the library does not require
the processors to share a common address space for their memory.
We therefore decided to require shared memory, a modern hardware
technology, which provides a common address space, and to use POSIX
threads to achieve parallelism.
\\
In 2D and 3D time evolutions of the conformal field equations the code
scales very well and on a 27 processor\footnote{The maximal number of
  processors available for a single user on our computer.} run we
typically get a speed-up of 24.
The remaining gap is mostly due to the variable work load per
gridpoint caused by the change of the equations near the boundary
as described in equation (I/19). 
For systems with a constant work load per gridpoint, e.~g.\ 3D Maxwell 
equations on periodic grids, we come significantly closer to the
optimal speed-up of 27.
\\
With the exception of dumping intermediate results to checkpoint, all
output of data is done in the XDR format to generate hardware independent
binaries.
The output interface allows the caller to request and get the output
of arbritrary rectangular sections of the grid with arbritrary
coarseness for each coordinate.
\\
The interaction between the parts of the code which provide the
computational infrastructure and the parts which are equation or
boundary specific has been minimised.
The code was successfully used to numerically solve initial value
problems for the 3D Maxwell equations on periodic grids, the 3D SU(2)
Yang-Mills equations on periodic grids, and the 1D, 2D, and 3D
conformal field equations for boundaries periodic in y and z
direction, for normal boundaries in all directions, and for octant
mode.
\\
In~\cite{Hu96mf} we describe why we believe that a numerical
relativity code should be able to deal with variables becoming
singular at gridpoints.
Since we want to also implement a similar strategy in more than 1D
later, the code is designed to allow a future extension to include
handling and bookkeeping of singular gridpoints.
Since the implementation of this extension was not necessary for the
purposes of this paper, it is due.
Its description will be the last part of this series.
%
%
%

\bibliography{biblio}
\bibliographystyle{prsty}

\end{document}